\documentclass[a4paper,UKenglish]{lipics}

\serieslogo{}
\volumeinfo
  {Carsten Fuhs}
  {1}
  {14th International Workshop on Termination 2014 (WST~'14)}
  {1}
  {1}
  {1}
\EventShortName{}

\def\copyrightline{}

\usepackage{microtype}
\usepackage{latexsym}
\usepackage{amsmath,amssymb,amstext,amsthm}
\usepackage{wasysym}

\usepackage{graphicx}
\usepackage{stmaryrd}

\usepackage{tikz}
\usetikzlibrary{arrows,automata,backgrounds,calc,chains,decorations.fractals,decorations.pathreplacing,fadings,fit,folding,mindmap,patterns,plotmarks,positioning,shadows,shapes.geometric,shapes.symbols,through,trees}

\usepackage{textcomp}

\renewcommand{\theenumi}{(\roman{enumi})}

\let\ocaption\caption
\renewcommand{\caption}[1]{\ocaption{\textit{#1}}}

\renewcommand{\mit}{\mathit}
\newcommand{\mrm}{\mathrm}

\newcommand{\msf}{\mathsf}

\newcommand{\mbb}{\mathbb}

\newcommand{\del}[1]{}
\newcommand{\sub}[2]{#1_{#2}}
\newcommand{\super}[2]{#1^{#2}}

\newcommand{\fap}[2]{#1(#2)}

\newcommand{\funnap}[3]{\fap{\super{#1}{#2}}{#3}}

\newcommand{\juxt}[2]{#1#2}

\newcommand{\relcomp}[2]{\mathrel{{#1}\cdot{#2}}}

\newcommand{\lam}[2]{\lambda#1.#2}
\newcommand{\mylam}{\lam}

\newcommand{\mmu}[2]{\mu#1.#2}

\newcommand{\infinity}{\infty}

\newcommand{\posemp}{\epsilon}

\newcommand{\ster}{\mit{Ter}}

\newcommand{\iter}{\fap{\ster^\infinity}}

\newcommand{\treeap}{\cdot}

\newcommand{\pairlft}{{\langle}}
\newcommand{\pairrgt}{{\rangle}}

\newcommand{\pairstr}[1]{\pairlft#1\pairrgt}

\newcommand{\tuple}{\pairstr}

\newcommand{\nat}{\mbb{N}}

\newcommand{\subst}[3]{#1[#2\,{{:}{=}\,}#3]}

\newcommand{\aclock}{%
\begin{tikzpicture}%
  \foreach \a in {0,30,...,330} {
    \draw [very thick] (\a:1.8mm) -- (\a:2.5mm);
  }
  \draw [thick] (0:0mm) -- (170:1.7mm);
  \draw [very thick] (0:0mm) -- (40:1.5mm);
\end{tikzpicture}%
}

\newcommand{\ctree}[1]{#1_{\hspace*{-.4ex}\smallclock}\hspace*{-.1ex}}

\newcommand{\sllt}{\msf{LLT}}

\newcommand{\sclevi}{\ctree{\sllt}}
\newcommand{\clevi}{\fap{\sclevi}}

\newcommand{\smred}{{\twoheadrightarrow}}
\newcommand{\mred}{\mathrel{\smred}}

\newcommand{\sconv}{\sub{{=}}{\beta}}
\newcommand{\conv}{\mathrel{\sconv}}
\newcommand{\notconv}{\not\conv}
\newcommand{\nconv}{\notconv}

\newcommand{\smhred}{{\twoheadrightarrow_\mit{h}}}
\newcommand{\mhred}{\mathrel{\smhred}}

\catcode`\@=11

\newcommand{\mywash}[2]{\setbox0=\hbox{$\m@th#1{#2}$}\wd0=0pt\box0}
\catcode`\@=12

\newcommand{\levy}{L\'{e}vy--Longo}
\newcommand{\levi}{\levy}

\newcommand{\boehm}{B\"{o}hm}
\newcommand{\bohm}{\boehm}

\newcommand{\sfpc}{\comb{Y}}
\newcommand{\fpc}[1]{\sub{\sfpc}{\!#1}}
\newcommand{\fpcC}{\fpc{0}}

\newcommand{\fpcT}{\fpc{1}}
\newcommand{\afpc}{Y}

\newcommand{\cxthole}{\hole}

\newcommand{\SNinf}{\mrm{SN}^{\infty}}

\newcommand{\UN}{\mrm{UN}}
\newcommand{\UNinf}{\UN^\infinity}
\newcommand{\CR}{\mrm{CR}}
\newcommand{\CRinf}{\CR^\infinity}

\newcommand{\app}{\juxt}
\newcommand{\leftappiterate}[3]{\app{#1}{\super{#2}{{\sim}#3}}}

\newcommand{\comb}{\msf}
\newcommand{\cS}{\comb{S}}

\newcommand{\cI}{\comb{I}}

\definecolor{MyRed}{rgb}{0.9,0.2,0.0}

\newcommand{\hole}{\raisebox{-2pt}{\scalebox{.7}[1.5]{$\Box$}}}

\renewcommand{\phi}{\varphi}

\newcommand{\dead}[1]{}

\newcommand{\old}[1]{\textcolor{Salmon}{.}}

\newcommand{\sred}{{\to}}
\newcommand{\red}{\mathrel{\sred}}

\def\ip#1#2{#1_{#2}}
\def\bitrm{\ip{\btrm}}

\newcommand{\sired}{\threeheadrightarrow}
\newcommand{\ired}{\mathrel{\sired}}

\newcommand{\mtsp}{\hspace{-1.665ex}}
\renewcommand{\twoheadrightarrow}{\mathrel{{\rightarrow\mtsp\rightarrow}}}
\newcommand{\threeheadrightarrow}{\mathrel{{\rightarrow\mtsp\rightarrow\mtsp\rightarrow}}}

\newcommand{\threeheadleftarrow}{\mathrel{{\leftarrow\mtsp\leftarrow\mtsp\leftarrow}}}

\newcommand{\sE}{\tau}
\newcommand{\E}{\fap{\sE}}
\newcommand{\nE}{\funnap{\sE}}

\newcommand{\sLL}{\smallclock}

\newcommand{\redLL}{\red_{\sLL}}
\newcommand{\siredLL}{\sired_{\sLL}}
\newcommand{\iredLL}{\ired_{\sLL}}

\newcommand{\smallclock}{\scalebox{.5}{\text{\aclock}}}

\newcommand{\Eter}{\iter{\lambda\sE}}

\newcommand{\sEp}[1]{\sE_{#1}}
\newcommand{\Ep}[1]{\fap{\sEp{#1}}}

\newcommand{\sEdel}{\sred_{\sE}}

\newcommand{\siEdel}{\sired_{\sE}}
\newcommand{\iEdel}{\mathrel{\siEdel}}

\newcommand{\synteq}{\equiv}

\newcommand{\btrm}{N}

\newcommand{\siredi}{{\threeheadleftarrow}}
\newcommand{\iredi}{\mathrel{\siredi}}

\newcommand{\pD}{\lambda}
\newcommand{\pL}{\mrm{L}}
\newcommand{\pR}{\mrm{R}}
\newcommand{\pT}{\sE}

\title{An Introduction to the Clocked Lambda Calculus\footnote{
  This paper has been published at the Workshop on Infinitary Rewriting 2014.
  It is a brief introduction to the work~\cite{endr:hend:klop:2010,endr:hend:klop:polo:2012,endrullis2013clocks,endr:hend:klop:polo:2013}.
}}
                             
\author[]{J\"{o}rg~Endrullis}
\author{Dimitri Hendriks}
\author{Jan Willem Klop}
\author{Andrew Polonsky}

\affil{VU University Amsterdam, The Netherlands}

\authorrunning{Endrullis, Hendriks, Klop and Polonsky}

\Copyright{Endrullis, Hendriks, Klop and Polonsky}

\subjclass{D.1.1, D.3.1, F.4.1, F.4.2, I.1.1, I.1.3}
\keywords{lambda calculus, convertibility, B\"ohm Trees}

\begin{document}

\maketitle

\begin{abstract}
  We give a brief introduction to the \emph{clocked $\lambda$-calculus},
  an extension of the classical $\lambda$-calculus with a unary symbol~$\sE$ 
  used to witness the $\beta$-steps.
  In contrast to the classical $\lambda$-calculus,
  this extension is infinitary strongly normalising and infinitary confluent.
  The infinitary normal forms are enriched \levi{} Trees, which we call \emph{clocked \levi{} Trees}.
\end{abstract}

\section{The Clocked Lambda Calculus}

The classical $\lambda$-calculus~\cite{bare:1984} is based on the $\beta$-rule
\begin{align*}
  \app{(\lam{x}{M})}{N}  \to \subst{M}{x}{N} 
\end{align*}
This calculus is neither infinitary normalising 
\begin{align*}
  (\lam{x}{xx})(\lam{x}{xx}) \to (\lam{x}{xx})(\lam{x}{xx}) \to \ldots \;,
\end{align*}
nor infinitary confluent.
To see this, let
\begin{align*}
  \fpcC \equiv \mylam{f}{\omega_f\omega_f} && \omega_f \equiv \mylam{x}{f(xx)}
\end{align*}
be Curry's fixed point combinator.
The term $\fpcC \cI$ admits the infinite (strongly convergent) rewrite sequences
\begin{align*}
  &\fpcC \cI \to_{\beta} (\lambda x. \cI(xx))(\lambda x. \cI(xx)) \ired \cI^{\omega} \\
  &\fpcC \cI \to_{\beta} (\lambda x. \cI(xx))(\lambda x. \cI(xx)) \to^2_{\beta} \Omega = (\lambda x.xx)(\lambda x.xx)
\end{align*}
Here infinitary confluence fails:
the terms $\cI^{\omega}$ and $\Omega$ have no common reduct since they reduce only to themselves
(see~\cite{bare:klop:2009} and~\cite[Chapter 12]{tere:2003}).
Even though infinitary confluence fails, the calculus has the property of infinitary unique normal forms.
When considering the $\beta$- and $\eta$-rule together, even this property fails,
see further~\cite{endr:hend:klop:2012,endr:grab:hend:klop:oost:2010}.

The \emph{clocked $\lambda$-calculus}~\cite{endr:hend:klop:polo:2013} consists of the following two rules:
\begin{align*}
  & \app{(\lam{x}{M})}{N}  \to \E{\subst{M}{x}{N}}\\
  & \app{\E{M}{N}} \to \E{\app{M}{N}}
\end{align*}
Here every $\beta$-step produces a symbol $\sE$ as a witness of the step.
The second rule is used to move the $\sE$'s out of the way of applications and hence potential $\beta$-redexes.
We write $\redLL$ for the reduction relation of the clocked $\lambda$-calculus.

For a simple example, consider the following reduction:
\begin{align*}
  \cI\cI\cI \redLL \E{\cI}\cI \redLL \E{\cI\cI} \redLL \E{\E{\cI}}
\end{align*}
where $\cI = \lam{x}{x}$.
\pagebreak
Note that the second step moves the $\sE$ out of the way of a $\beta$-redex.

As a second example, let us consider Curry's fixed point combinator:
\begin{align*}
  \app{\fpcC}{f} 
  & \equiv \app{(\lam{f}{\app{\omega_f}{\omega_f}})}{f} 
    \redLL \E{\app{\omega_f}{\omega_f}}  \\
  \app{\omega_f}{\omega_f} 
  & \redLL \E{\app{f}{(\app{\omega_f}{\omega_f}})}
\end{align*}  
Hence $\app{\fpcC}{f}$ rewrites to the infinite normal form 
\begin{align*}
  \app{\fpcC}{f} \iredLL \E{\E{f(\E{f(\E{f(\ldots)})})}}
\end{align*}
written without brackets as $\sE\sE z\sE z\sE z\ldots$.

The clocked $\lambda$-calculus enjoys the properties of infinitary confluence, infinitary strong normalization~\cite{klop:vrij:2005,zant:2008,endr:grab:hend:klop:vrij:2009}
and hence infinitary unique normal forms:
\begin{description}\setlength{\itemsep}{.55ex}
\vspace*{0.55ex}
\item 
  [$\SNinf$]: all infinite rewrite sequences are strongly convergent;
\item
  [$\CRinf$]: 
  \(\forall M,\bitrm{1}, \bitrm{2}\;
       (
       \bitrm{1} \iredi_R M \ired_R \bitrm{2}
         \;\implies\;
       \bitrm{1} \relcomp{\ired_R}{\iredi_R} \bitrm{2}
       )
  \);
\item
  [$\UNinf$]:
  \( 
    \forall M,\bitrm{1}, \bitrm{2}\;
      ( 
      \bitrm{1} \iredi_R M \ired_R \bitrm{2}
        \text{ and } 
        \text{$\bitrm{1},\, \bitrm{2}$ normal forms}  
          \;\implies\;
            \bitrm{1} \synteq \bitrm{2}
            )  
  \).
\end{description}

\begin{lemma}\label{lem:SNinf:UNinf:CRinf}
  The 
  relation $\siredLL$ 
  has the properties $\CRinf$, $\SNinf$ and $\UNinf$.
\end{lemma}

\section{Clocked \levi{} Trees}

The unique infinitary normal forms with respect to $\siredLL$ 
are \emph{clocked \levi{} Trees}~\cite{endr:hend:klop:2010,endr:hend:klop:polo:2012,endr:hend:klop:polo:2013}, that is,
\levi{} Trees (a variant of \boehm{} Trees) enriched with symbols $\sE$
witnessing the $\beta$-steps performed in the reduction to the normal form.
We write $\clevi{M}$ for the unique infinite normal form of $M$.

Consider the well-known fixed point combinators 
of Curry and Turing, $\fpcC$ and $\fpcT$:
\begin{align*}
  \fpcC & \equiv \lam{f}{\omega_f\omega_f} &
  \fpcT & \equiv \eta \eta
  \\
  \omega_f & \equiv \lam{x}{f(xx)} &
  \eta & \equiv \lam{xf}{f (xxf)}
\end{align*}
Figure~\ref{fig:boem:y0:y1} displays the clocked \levi{} Trees of $\fpcC f$ (left) and $\fpcT f$ (right)
where we write $\sE^n(t)$ for $\underbrace{\sE(\sE(\ldots(\sE(}_{\text{$n$-times}}t))\ldots))$.
\begin{figure}[ht!]
  \centering
  \begin{tikzpicture}[level distance=7mm,inner sep=1mm]
\node (sE^2) {$\sE^2$}
    child { node (treeap) {$\treeap$}
      child { node (f) {$f$}
    }
      child { node (sE^1) {$\sE^1$}
        child { node {$\treeap$}
          child { node (f) {$f$}
        }
          child { node {$\sE^1$}
            child { node {$\treeap$}
              child { node (f) {$f$}
            }
              child { node (ddots) {$\ddots$}
            }
          }
        }
      }
    }
  };
\begin{pgfonlayer}{background}
\end{pgfonlayer}

  \end{tikzpicture}
  \begin{tikzpicture}[level distance=7mm,inner sep=1mm]
\node (sE^2) {$\sE^2$}
    child { node (treeap) {$\treeap$}
      child { node (f) {$f$}
    }
      child { node {$\sE^2$}
        child { node {$\treeap$}
          child { node (f) {$f$}
        }
          child { node {$\sE^2$}
            child { node {$\treeap$}
              child { node (f) {$f$}
            }
              child { node (ddots) {$\ddots$}
            }
          }
        }
      }
    }
  };
\begin{pgfonlayer}{background}
\end{pgfonlayer}

  \end{tikzpicture}
  \caption{Clocked \levi{} Trees $\clevi{\fpcC f}$ and $\clevi{\fpcT f}$ of $\fpcC f$ and $\fpcT f$, respectively.}
  \label{fig:boem:y0:y1}
\end{figure}
For $\fpcC f$ we have seen the reduction to the infinite normal form above,
and a similar computation leads to the clocked \levi{} Tree of $\fpcT f$.
The $\sE$'s in the clocked \levi{} Tree witness the number of head reduction steps needed to
normalise the corresponding subterm to weak head normal form.
 
\section{Discriminating Lambda Terms}

For more details on the results in this section, we refer to~\cite{endr:hend:klop:polo:2013,endr:hend:klop:polo:2012}.

We define ${\sEdel}$ 
by the rule
\begin{align*}
  \E{M} \to M
\end{align*}
and use $=_\sE$ to denote the equivalence closure of $\sEdel$.
For $M,N \in \Eter$, we define
\begin{enumerate}
  \item 
    $M \succeq_{\smallclock} N$, \emph{$M$ is globally improved by $N$} 
    iff $\clevi{M} \iEdel \clevi{N}$;
  \item 
    $M =_{\smallclock\exists} N$, \emph{$M$ eventually matches $N$} 
    iff $\clevi{M} =_{\sE} \clevi{N}$.
\end{enumerate}

For example, $\fpcC f$ globally improves $\fpcT f$ ($\fpcC f \preceq_{\smallclock} \fpcT f$)
as can be seen from the clocked \levi{} Trees 
of $\fpcC f$ and $\fpcT f$ in Figure~\ref{fig:boem:y0:y1}.

\begin{theorem}
  Clocks are accelerated under reduction, 
  that is, if $M {\mred} N$, then the reduct $N$ improves $M$ globally,
  that is,
  $\clevi{M} \iEdel \clevi{N}$.
\end{theorem}

As a consequence we obtain the following method for discriminating $\lambda$-terms:
\begin{theorem}\label{thm:general}
  Let $M$ and $N$ be $\lambda$-terms.
  If $N$ cannot be improved 
  globally by any reduct of $M$, then $M \ne_\beta N$.
\end{theorem}

In~\cite{endr:hend:klop:polo:2012} we use this theorem to answer the following 
question of Selinger and Plotkin~\cite{plot:2007}:
\textit{Is there a fixed point combinator $\afpc$ such that}
\begin{align*}
  A_\afpc \equiv \afpc(\mylam{z}{fzz}) \conv \afpc(\mylam{x}{\afpc(\mylam{y}{fxy})}) \equiv B_\afpc
  \label{eq:plotkin}
\end{align*}
or in other notation: 
\begin{align*}
  \mmu{z}{fzz} \conv \mmu{x}{\mmu{y}{fxy}}\;,
\end{align*}
with the usual definition $\mmu{x}{M(x)} = Y(\mylam{x}{M(x)})$.
The terms $A_\afpc$ and $B_\afpc$ have the same \bohm{} Trees, namely
the solution of $T = f T T$.
Clocked \levi{} Trees can be employed to show
that such fixed point combinators do not exist, see~\cite{endr:hend:klop:polo:2012}.
For deciding equality of $\mu$-terms with the usual unfolding rule $\mmu{z}{M(z)} = \subst{M}{z}{\mmu{z}{M(z)}}$, see~\cite{endr:grab:klop:oost:2011}.

For a large class of $\lambda$-terms the clocks are invariant under reduction,
that is, the clocked \levi{} Trees coincide up to insertion and removal of a finite number of~$\sE$'s.
\begin{definition}[Simple terms]
  A redex $(\lam{x}{M}){N}$ is called:
  \begin{enumerate}\setlength{\itemsep}{0ex}
    \item \emph{linear} if $x$ has at most one occurrence in $M$;
    \item \emph{call-by-value} if $N$ is a normal form; and
    \item \emph{simple} if it is linear or call-by-value.
  \end{enumerate}
  A $\lambda$-term $M$ is \emph{simple}
  if (a) it has no weak head normal form,
  or the head reduction to whnf contracts only simple redexes
  and is of one of the following forms: 
  (b) $M \mhred \lam{x}{M'}$ with $M'$ a simple term, or
  (c) $M \mhred {y M_1 \ldots M_m}$ with $M_1,\ldots,M_m$ simple terms.
\end{definition}
Note that this definition is inherently coinductive;
this is similar to the definition of \boehm{} Trees in~\cite{bare:1984}.
The infinitary rewrite relation itself can also be defined coinductively,
see further~\cite{coqu:1996,endr:polo:2012b,DBLP:journals/corr/EndrullisHHPS13}.

\begin{theorem}\label{prop:simple}
  Let $N$ be a reduct of a simple term $M$. Then $N$ eventually matches~$M$
  (i.e., $\clevi{M} =_{\sE} \clevi{N}$).
\end{theorem}
For simple terms, the discrimination method can be simplified as follows:
\begin{theorem}\label{cor:simple:simple}
  If simple terms $M$, $N$ do not eventually match ($\clevi{M} \not=_{\sE} \clevi{N}$),
  then they are not $\beta$-convertible, that is, $M \ne_\beta N$.
\end{theorem}

\begin{example}\label{ex:boehm:seq}
  We show that the fixed point combinators  
  $\fpc{0},\fpc{1},\fpc{2},\ldots$ of the \boehm{} sequence
  are all inconvertible.
  For $n \ge 1$, define
  \begin{align*}
    \fpc{n} = \leftappiterate{\eta\eta}{\delta}{n-1}
  \end{align*} 
  where
  \begin{align*}
    & \leftappiterate{M}{N}{0} = M \\
    & \leftappiterate{M}{N}{n+1} = \leftappiterate{MN}{N}{n}
  \end{align*}
  The clocked \levi{} Trees of $\fpc{0} x$ and $\fpc{1} x$ are shown in Figure~\ref{fig:boem:y0:y1}.
  We now determine the clocked \levi{} Trees of $\fpc{n} x$ for $n \geq 2$:
  \begin{align*}
    \fpc{n}
    & \equiv
    \leftappiterate{\eta \eta}{\delta}{n-1} x \\
    & \redLL \leftappiterate{\E{\lam{f}{f(\eta\eta f)}}}{\delta}{n-1} x \\
    & \redLL^* \E{ \leftappiterate{(\lam{f}{f(\eta\eta f)})\delta}{\delta}{n-2} x } \\
    & \redLL \E{ \leftappiterate{\E{\delta(\eta\eta \delta)}}{\delta}{n-2} x } \\
    & \redLL^* \nE{2}{ \leftappiterate{\delta(\eta\eta \delta)}{\delta}{n-2} x } \\
    & \redLL^* \nE{4}{ \leftappiterate{\delta(\eta\eta \delta\delta)}{\delta}{n-3} x } \\
    &\vdots\\
    & \redLL^* \nE{2n-2}{ \delta(\eta\eta \leftappiterate{}{\delta}{n-1}) x } \\
    & \redLL^* \nE{2n}{ x(\eta\eta \leftappiterate{}{\delta}{n-1} x)}
  \end{align*}
  None of these steps duplicate a redex, 
  hence $\fpc{n}$ is a simple term. 
  We have
  \begin{align*}
    \clevi{\fpc{n} x} \equiv \nE{2n}{x \;\clevi{\fpc{n} x}}
  \end{align*}
  Observe that all of the clocked \levi{} Trees $\clevi{\fpc{n} x}$
  differ in an infinite number of~$\sE$'s.
  By Theorem~\ref{cor:simple:simple}
  it follows that all terms in the \boehm{} sequence are inconvertible.
\end{example}

\section{Atomic Clocked Lambda Calculus}

The clocked $\lambda$-calculus can be enhanced
to not only recording whether head reduction steps have taken place,
but also where they took place.
We use $\{\pD,\pL,\pR,\pT\}^*$ for the positions.

The \emph{atomic clocked $\lambda$-calculus} consists of the rules
  \begin{align*}
    \app{(\lam{x}{M})}{N}
    & \to \Ep{\posemp}{\subst{M}{x}{N}} \\
    \app{\Ep{p}{M}}{N}
    & \to \Ep{\pL p}{\app{M}{N}}
  \end{align*}
The atomic clocks further strengthen the discrimination power of method \levi{} Trees.

Let $\comb{S} = \lam{abc}{ac(bc)}$.
For $k, n_1,\ldots,n_k \in \nat$ define a fixed point combinator $\comb{Y}^{\tuple{n_1,\ldots,n_k}}$ by
\begin{align*}
  \comb{Y}^{\tuple{n_1,\ldots,n_k}} = \comb{G}_{n_k}[\ldots \comb{G}_{n_1}[\fpcC]\ldots]
\end{align*}
where $\comb{G}_n = \leftappiterate{\cxthole(\cS\cS)}{\cS}{n}\cI$.

As fixed point combinators, they all have the same \levi{} Tree $\lam{x}{x(x(x(\ldots)))}$. 
However, using atomic clocked \levi{} Trees
we have shown in~\cite{endr:hend:klop:polo:2012} 
that all these fixed point combinators are different, all of them are inconvertible:
$\vec{n} \ne \vec{m}$ implies $\comb{Y}^{\vec{n}} \nconv \comb{Y}^{\vec{m}}$.

%

\section{Future Work}

We have employed the (atomic) clocked $\lambda$-calculus for proving that $\lambda$-terms
are not convertible by showing that they have a different tempo 
in reducing to their infinite normal form.
The method is however not yet strong enough to answer questions like:
\emph{Is there a fixed point combinator $Y$ such that}
\begin{align*}
  Y & \conv \delta\, Y \\
  Y & \conv Y \delta 
\end{align*}
where $\delta = \lam{ab}{b(ab)}$?
R.~Statman conjectured that no such fixed point combinator exists.
However, this is still an open problem\footnote{
  B.~Intrigila~\cite{intr:1997} gave a proof that no such fixed point combinator exists. 
  The proof however contains a serious gap, see further~\cite{endr:hend:klop:polo:2013}.
}.
It would be interesting to see whether methods in the flavour of the
clocked $\lambda$-calculus could contribute to a solution.
Note that every fixed point combinator fulfils the first equation:
$Y = \delta \, Y$ if and only if $Y$ is a fixed point combinator, that is, 
all fixed point combinators are fixed points of $\delta$.

Furthermore, we are interested to investigate further applications of
the clocked $\lambda$-calculus.
For example, the clocks can be used as a measure of efficiency.

\bibliography{main.bib}

\end{document}